\documentclass[twocolumn,numerical,superscriptaddress,amsmath,amssymb,aip,apl]{revtex4}
\usepackage{graphicx}
\usepackage{dcolumn}
\usepackage{bm}

\usepackage{epsf}
\usepackage{epsfig}
\usepackage[latin1]{inputenc}   

\usepackage{amsmath}
\usepackage{amsthm}
\usepackage{amssymb}
\usepackage{epsfig}
\usepackage[bf,footnotesize]{caption}

\begin{document}

\title{Defect evolution and interplay in n-type InN}
\author{Christian Rauch}
\email{christian.rauch@aalto.fi} \affiliation{Department of Applied
Physics, Aalto University, P.O. Box 11100, FI-00076 Aalto, Espoo,
Finland}
\author{Filip Tuomisto}
\affiliation{Department of Applied Physics, Aalto University, P.O.
Box 11100, FI-00076 Aalto, Espoo, Finland}
\author{Arantxa Vilalta-Clemente}
\affiliation{CIMAP UMR 6252 CNRS-ENSICAEN-CEA-UCBN, 6, Boulevard du
Mar\'{e}chal Juin, 14050 Caen Cedex, France}
\author{Bertrand Lacroix}
\affiliation{CIMAP UMR 6252 CNRS-ENSICAEN-CEA-UCBN, 6, Boulevard du
Mar\'{e}chal Juin, 14050 Caen Cedex, France}
\author{Pierre Ruterana}
\affiliation{CIMAP UMR 6252 CNRS-ENSICAEN-CEA-UCBN, 6, Boulevard du
Mar\'{e}chal Juin, 14050 Caen Cedex, France}
\author{Simon Kraeusel}
\affiliation{Department of Physics, SUPA, Strathclyde University, G4
0NG Glasgow, United Kingdom}
\author{Ben Hourahine}
\affiliation{Department of Physics, SUPA, Strathclyde University, G4
0NG Glasgow, United Kingdom}
\author{William J. Schaff}
\affiliation{Department of Electrical and Computer Engineering,
Cornell University, 425 Philips Hall, Ithaca, New York 14853, USA}
\date{\today}

\begin{abstract}
The nature and interplay of intrinsic point and extended defects in
n-type Si-doped InN epilayers with free carrier concentrations up to
$6.6 \times 10^{20}$cm$^{-3}$ are studied using positron
annihilation spectroscopy and transmission electron microscopy and
compared to results from undoped irradiated films. In as-grown
Si-doped samples, $V_{\text{In}}$-$V_{\text{N}}$ complexes are the
dominant III-sublattice related vacancy defects. Enhanced formation
of larger $V_{\text{In}}$-$mV_{\text{N}}$ clusters is observed at
the interface, which speaks for a high concentration of additional
$V_{\text{N}}$ in the near-interface region and coincides with an
increase of the dislocation density in that area.
\end{abstract}

\maketitle

InN possesses a strong propensity for n-type conductivity which can
be explained by an exceptionally high Fermi stabilization
energy~\cite{King2008} well above the conduction band minimum.
Taming the conductivity is one requirement for exploiting the
material's high potential for electronic and opto-electronic
devices~\cite{Monemar1999}. Therefore, a deep understanding of the
defect landscape in n-type InN is required. \textit{Ab-initio}
calculations predict that hydrogen acts as an effective donor
impurity in InN~\cite{Janotti2008}, while $V_{\text{N}}$ and
$V_{\text{In}}$ should be the dominant intrinsic donor and acceptor
type point defects~\cite{Stampfl2000}. Additionally, high densities
of extended defects are commonly found in as-grown material and have
been correlated with an electron accumulation layer at InN
interfaces~\cite{Piper2006}. In this letter, we use positron
annihilation spectroscopy (PAS) and transmission electron microscopy
(TEM) to study the evolution and interplay of native point and
extended defects in highly n-type InN under different conditions.
Si-doped InN layers~\cite{Schaff2004} with free electron
concentrations from $4.5 \times 10^{19}$~-~$6.6 \times
10^{20}$cm$^{-3}$ are investigated and compared to results from an
undoped ($n_{e}=1\times10^{18}$cm$^{-3}$ before irr.) irradiated InN
film~\cite{Jones2007} before ($3.2 \times 10^{20}$cm$^{-3}$) and
after annealing ($6 \times 10^{19}$cm$^{-3}$). All films were
deposited by plasma-assisted molecular beam epitaxy (PAMBE) as
$\sim$500nm thick layers on c-plane sapphire substrates with a GaN
buffer
layer~\cite{Jones2007,Schaff2004}.\\
\begin{figure} \centering
\includegraphics[width=0.9\linewidth]{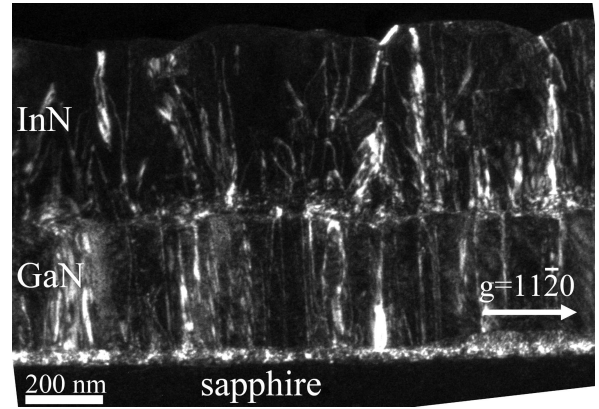}
  \caption{Cross-sectional dark-field TEM micrograph (g=11-20) of a representative Si-doped sample (n$_{e}$=4.0$\times$$10^{20}$~cm$^{-3}$), showing edge and mixed type dislocations.}
  \label{TEM}
\end{figure}
TEM measurements of thin cross-sectional samples were performed
using a JEOL 2010 operating at an acceleration voltage of
200~kV~\cite{Dasilva2010}. Fig.~\ref{TEM} shows a TEM micrograph
obtained in weak beam (WB) conditions with g=11-20 for a
representative Si-doped sample. Edge and mixed type dislocations are
visible distributed throughout the InN layer with an average density
of $4.0 \times 10^{9}$cm$^{-2}$ and $1.1 \times 10^{9}$cm$^{-2}$,
respectively. An agglomeration of dislocations close to the InN/GaN
interface can be noticed.  The density of screw type dislocations is
$3.1 \times 10^{8}$cm$^{-2}$ which corresponds to $\sim$6\% of the
total dislocation density. Additionally, a high density ($3 \times
10^{5}$cm$^{-1}$) of stacking faults was revealed for WB
conditions~\cite{Dasilva2010} with g=10-10 (not shown here). In the
irradiated InN film, earlier TEM results~\cite{LilientalWeber2007}
showed irradiation-induced formation dislocation loops in addition
to a significant density of planar defects introduced during growth.
After annealing at 475 $^{\circ}$C the density of dislocation loops
increased from $2.2 \times 10^{10}$cm$^{-2}$ to $9.0 \times
10^{10}$cm$^{-2}$~\cite{LilientalWeber2007,Jones2007}. Vacancy
agglomeration after annealing was proposed as reason for this
increase.\\
We applied PAS to investigate vacancy-type point defects and their
nature in the InN samples. Using a mono-energetic positron beam,
depth-dependent Doppler broadening spectra were recorded at room
temperature to probe the momentum distribution of annihilating
electron-positron pairs. Details on the experimental technique and
setup can be found elsewhere~\cite{Saarinen1998,Rauch2011c}.
\begin{figure} [t] \centering
\includegraphics[width=1\linewidth]{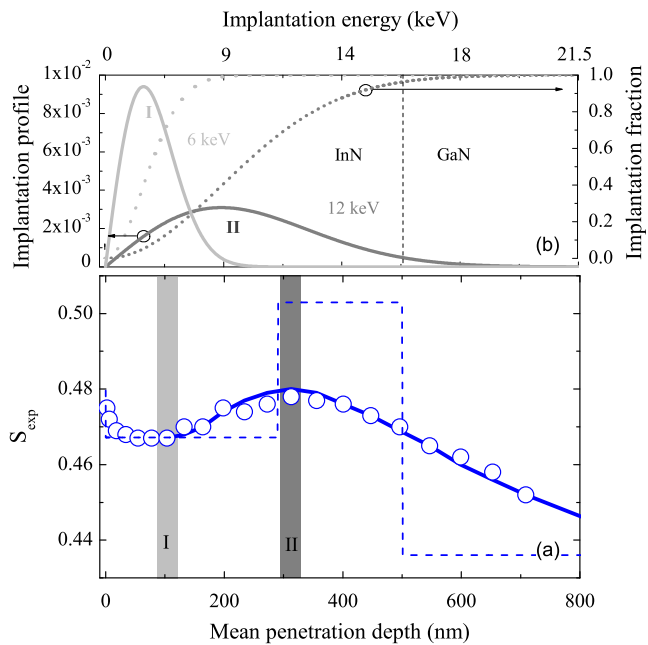}
  \caption{(Color online) Measured S parameter (open circles) of sample 2 as function of the mean positron implantation depth (a). Corresponding implantation energies are given for comparison. The solid line shows a fit of the data using a simple three-layer model of the S-parameter (dashed line). Fig.~(b) shows the calculated positron implantation profiles and fractions for positron implantation energies of 6~keV (I) and 12~keV (II).}
  \label{ImplantationProfile}
\end{figure}
Fig.~\ref{ImplantationProfile}(a) shows the measured S-parameters of
a representative Si-doped sample for positron implantation energies
from 0-20~keV. After annihilation at surface-specific states for low
implantation energies, the S-parameter drops quickly to a local
minimum at $\sim$6 keV. Comparison with the positron implantation
profile at that energy [Fig.~\ref{ImplantationProfile}(b)] reveals
that this point is representative for annihilations from the first
150nm of the sample, with a mean implantation depth of
$\bar{x}$=100nm. Deeper inside the sample, the S-parameter increases
to a local maximum at $\sim$12 keV (corresponding to a mean
implantation depth of $\bar{x}$=310nm) and positrons probe a wide
region reaching the interface to the GaN buffer layer. For higher
implantation energies a significant amount of positrons annihilate
in the GaN buffer layer pulling the measured S-parameter towards the
value of the GaN lattice. The solid curve in
Fig.~\ref{ImplantationProfile}(a) shows a fit of the measured
spectrum using the multi-layer fitting program
VEPFIT~\cite{Veen1995}. It reveals that the experimental spectrum
can be well described assuming a two-layer structure of the
S-parameter inside the InN film (see dashed line) with a 300~nm
thick near-surface and 200~nm thick near-interface layer and a
positron diffusion length of $\sim$5~nm.\\
\begin{figure} [t]
\centering
\includegraphics[width=0.8\linewidth]{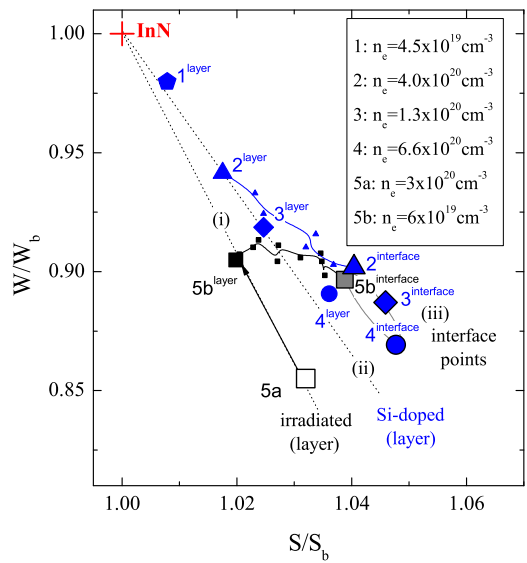}
  \caption{(Color online) S and W line-shape parameters of the Doppler broadened electron-positron annihilation $\gamma$-radiation for different samples and areas.}
  \label{SW}
\end{figure}
Representative for the near-surface ("layer") and near-interface
("interface") areas, the measured S and W parameter at 6 and 12~keV
are plotted in Fig.~\ref{SW} together with accordingly determined
values of the remaining samples. All points are normalized by the
value of an InN reference sample for which no positron trapping at
open volume defects is observed~\cite{Rauch2011c}. Samples 1 and 5a
did not exhibit any depth-profile and therefore only one set of
parameters is displayed. For all Si-doped samples (samples 1-4), the
"layer" points fall on one line through the reference value of the
InN lattice. Hence~\cite{Saarinen1998}, one dominant vacancy-type
positron trap is present in this area with increasing
(room-temperature) annihilation fraction from sample 1-4. For the
"interface" points, a deviation from the "layer" line is visible due
to increasing S-parameters and comparably less pronounced decrease
in the W-parameters. Therefore, a different dominant vacancy-type
positron trap has to be expected here. The slope defined by the
irradiated sample before annealing (sample 5a) is
steeper~\cite{Tuomisto2007e} than for the as-grown samples. Upon
rapid thermal annealing (RTA)~\cite{Jones2007} (sample 5b), a
profile in the depth dependent spectrum of the S-parameter is
developed~\cite{Reurings2010} with a layer and interface-specific
value. The near-surface "layer"-point is shifted closer toward the
InN lattice point but remains on the same line as the as-irradiated
sample. This indicates a decrease in annihilation fraction at the
same positron trap as before annealing. The "interface" point,
however, deviates strongly after annealing and is moved close to the
interface points of the Si-doped samples.\\
We find 3 different dominant vacancy-type positron traps in the InN
samples, i.e., defects created by high-energy particle irradiation
(i), defects dominant at the near-surface area of as-grown Si-doped
samples (ii), and defects responsible for the observed changes at
the interfaces of both Si-doped as well as RTA-treated, irradiated
samples (iii). A comparison of high-resolution coincidence Doppler
broadening spectra with density functional theory (DFT) calculations
of positron trapping and annihilation in InN
reveals~\cite{Rauch2011c} that these positron traps can be
identified as (i) isolated In vacancies ($V_{\text{In}}$), (ii)
mixed In-N divacancies ($V_{\text{In}}$-$V_{\text{N}}$), and (iii)
bigger $V_{\text{In}}$-$mV_{\text{N}}$ ($m\approx2,3$) vacancy
complexes, respectively. High-energy particle irradiation introduces
isolated $V_{\text{In}}$ as dominant vacancy-type positron traps in
InN. Subsequent annealing leads to a re-arrangement of vacancy
defects~\cite{Reurings2010}, as observable in both TEM and positron
annihilation measurements. $V_{\text{In}}$ become mobile at or below
the annealing temperature and start to move toward the surface and
the interface with the GaN buffer, respectively, where they either
recombine, anneal out (at the surface) or form complexes with
residual $V_{\text{N}}$ (interface). Based on the employed annealing
temperature of 475$^{\circ}$C, we can
estimate~\cite{Limpijumnong2004} an upper limit of $E_{b}\leq
1.9~eV$ for the migration barrier of the $V_{\text{In}}$. This is in
good agreement with the calculated value of 1.6~eV~\cite{Duan2009a}
and indicates that isolated $V_{\text{In}}$ are mobile during InN
growth (assuming usual growth temperatures of $\sim$~$550^{\circ}$C
for, e.g., MBE).\\ No isolated $V_{\text{In}}$ are observed in our
measurements of as-grown Si-doped InN. Instead, we find
$V_{\text{In}}$-$V_{\text{N}}$ complexes. Hence, we conclude that
in-grown $V_{\text{In}}$ are stabilized through the formation of
complexes with $V_{\text{N}}$. This is supported by recent DFT
results~\cite{Duan2009a} which predict a positive binding energy
between $V_{\text{In}}$ and $V_{\text{N}}$. Vacancy-stabilization
through the formation of vacancy-donor complexes has been observed
also in GaN (Ref.~\cite{Tuomisto2007c}, and references therein) and
AlN~\cite{Maki2011}. The increased incorporation of $V_{\text{In}}$
complexes with increasing free electron concentration suggests
strongly that $V_{\text{In}}$-related defects act as a source of
compensation in n-type InN, which is in line with theoretical
results~\cite{Stampfl2000}. The enhanced formation of larger
$V_{\text{In}}$-$mV_{\text{N}}$ complexes toward the interface with
the GaN buffer layer (in irradiated material after annealing as well
as Si-doped samples) indicates that the InN/GaN interface is
attractive for vacancy defects. An additional high density of
$V_{\text{N}}$ in that area could provide the proximity required for
the promotion of efficient vacancy $V_{\text{In}}$-$mV_{\text{N}}$
clustering. However, neutral and positively charged isolated
$V_{\text{N}}$ and $mV_{\text{N}}$-complexes can not be detected in
PAS measurements~\cite{Rauch2011c}. Duan \textit{et al.} have
calculated~\cite{Duan2008} a positive binding energy between single
$V_{\text{N}}$ under n-type conditions and a strong tendency for the
formation of larger $V_{\text{N}}$ clusters. Hence, the formation of
$V_{\text{In}}$-$mV_{\text{N}}$ complexes could occur through a
precursor state of $mV_{\text{N}}+ V_{\text{In}}$ $\rightarrow$
$V_{\text{In}}$-$mV_{\text{N}}$, in accordance to what has been
proposed earlier in Mg-doped InN~\cite{Uedono2009b}.\\
Based on the TEM data, the observed increase in vacancy clustering
at the InN/Gan interface coincides with elevated dislocation
densities in that area. In order to assess the effect of
dislocations on the formation energies of point defects in their
vicinity we performed DFT calculations of strained InN lattices. We
found that typical strain associated with screw dislocations (0-15\%
shear) decreases the formation energies of $V_{\text{In}}$ and
$V_{\text{N}}$ only slightly by $\leq$30~meV, and hence should not
play any major role. Investigations on the effects of edge
dislocations are under way. Besides strain-related influences on the
defect formation energies, additional dislocation-related vacancy
formation mechanisms such as dislocation movement and/or decoration
of dislocations might be possible. In GaN, recent theoretical
calculations~\cite{Kraeusel2011} suggest stable configurations of
vacancies inside dislocation cores and a correlation between vacancy
densities and dislocations was found~\cite{Oila2003}. It should be
noted that dislocations might also directly affect the positron
annihilation signal~\cite{Tengborn2006} by, e.g., forming shallow
traps for positrons. The exceptionally low values of the positron
diffusion length in the InN samples do support the presence of such
additional positron trapping centers with annihilation
characteristics close to the bulk. Detailed theoretical
investigations on positron trapping and annihilation at dislocations
in wurtzite semiconductors are currently being performed.\\
In summary, combining results from PAS and TEM we find that isolated
$V_{\text{In}}$ are only present in irradiated InN films and anneal
out at temperatures of $\leq475$$^{\circ}$C if not stabilized by
other point defects. Stabilization of $V_{\text{In}}$ occurs through
complex formation with $V_{\text{N}}$.
$V_{\text{In}}$-$mV_{\text{N}}$ complexes are the dominant
vacancy-type positron trap in as-grown InN samples. Toward the
interface between the InN layer and the GaN buffer, enhanced
formation of bigger vacancy clusters with increasing number of
$V_{\text{N}}$ is observed in both as-grown and irradiated material
after annealing and coincides with increased dislocation densities
in that area. This indicates that the InN/GaN interface is strongly
attractive for vacancy defects and points at elevated concentrations
of additional $V_{\text{N}}$ and
$V_{\text{N}}$-complexes in that area.\\
The authors wish to thank T. Veal for helpful discussions. This work
has been supported by the European Commission under the 7th
Framework Program through the Marie Curie Initial Training Network
RAINBOW, Contract No. PITN-Ga-2008-213238, and the Academy of
Finland.


\begin{thebibliography}{23}
\bibliographystyle{apsrev}
\expandafter\ifx\csname
natexlab\endcsname\relax\def\natexlab#1{#1}\fi
\expandafter\ifx\csname bibnamefont\endcsname\relax
  \def\bibnamefont#1{#1}\fi
\expandafter\ifx\csname bibfnamefont\endcsname\relax
  \def\bibfnamefont#1{#1}\fi
\expandafter\ifx\csname citenamefont\endcsname\relax
  \def\citenamefont#1{#1}\fi
\expandafter\ifx\csname url\endcsname\relax
  \def\url#1{\texttt{#1}}\fi
\expandafter\ifx\csname urlprefix\endcsname\relax\def\urlprefix{URL
}\fi \providecommand{\bibinfo}[2]{#2}
\providecommand{\eprint}[2][]{\url{#2}}

\bibitem[{\citenamefont{King et~al.}(2008)\citenamefont{King, Veal, Jefferson,
  Hatfield, Piper, McConville, Fuchs, Furthm\"{u}ller, Bechstedt, Lu
  et~al.}}]{King2008}
\bibinfo{author}{\bibfnamefont{P.~D.~C.} \bibnamefont{King}},
  \bibinfo{author}{\bibfnamefont{T.~D.} \bibnamefont{Veal}},
  \bibinfo{author}{\bibfnamefont{P.~H.} \bibnamefont{Jefferson}},
  \bibinfo{author}{\bibfnamefont{S.~A.} \bibnamefont{Hatfield}},
  \bibinfo{author}{\bibfnamefont{L.~F.~J.} \bibnamefont{Piper}},
  \bibinfo{author}{\bibfnamefont{C.~F.} \bibnamefont{McConville}},
  \bibinfo{author}{\bibfnamefont{F.}~\bibnamefont{Fuchs}},
  \bibinfo{author}{\bibfnamefont{J.}~\bibnamefont{Furthm\"{u}ller}},
  \bibinfo{author}{\bibfnamefont{F.}~\bibnamefont{Bechstedt}},
  \bibinfo{author}{\bibfnamefont{H.}~\bibnamefont{Lu}}, \bibnamefont{et~al.},
  \bibinfo{journal}{Phys. Rev. B} \textbf{\bibinfo{volume}{77}},
  \bibinfo{pages}{045316} (\bibinfo{year}{2008}).

\bibitem[{\citenamefont{Monemar}(1999)}]{Monemar1999}
\bibinfo{author}{\bibfnamefont{B.}~\bibnamefont{Monemar}}, \bibinfo{journal}{J.
  Mater. Sci. Mater. Electron.} \textbf{\bibinfo{volume}{10}},
  \bibinfo{pages}{227} (\bibinfo{year}{1999}).

\bibitem[{\citenamefont{Janotti and de~Walle}(2008)}]{Janotti2008}
\bibinfo{author}{\bibfnamefont{A.}~\bibnamefont{Janotti}} \bibnamefont{and}
  \bibinfo{author}{\bibfnamefont{C.~G.~Van} \bibnamefont{de~Walle}},
  \textbf{\bibinfo{volume}{92}}, \bibinfo{pages}{032104}
  (\bibinfo{year}{2008}).

\bibitem[{\citenamefont{Stampfl et~al.}(2000)\citenamefont{Stampfl, Van~de
  Walle, Vogel, Kr\"uger, and Pollmann}}]{Stampfl2000}
\bibinfo{author}{\bibfnamefont{C.}~\bibnamefont{Stampfl}},
  \bibinfo{author}{\bibfnamefont{C.~G.} \bibnamefont{Van~de Walle}},
  \bibinfo{author}{\bibfnamefont{D.}~\bibnamefont{Vogel}},
  \bibinfo{author}{\bibfnamefont{P.}~\bibnamefont{Kr\"uger}}, \bibnamefont{and}
  \bibinfo{author}{\bibfnamefont{J.}~\bibnamefont{Pollmann}},
  \bibinfo{journal}{Phys. Rev. B} \textbf{\bibinfo{volume}{61}},
  \bibinfo{pages}{R7846} (\bibinfo{year}{2000}).

\bibitem[{\citenamefont{Piper et~al.}(2006)\citenamefont{Piper, Veal,
  McConville, Lu, and Schaff}}]{Piper2006}
\bibinfo{author}{\bibfnamefont{L.~F.~J.} \bibnamefont{Piper}},
  \bibinfo{author}{\bibfnamefont{T.~D.} \bibnamefont{Veal}},
  \bibinfo{author}{\bibfnamefont{C.~F.} \bibnamefont{McConville}},
  \bibinfo{author}{\bibfnamefont{H.}~\bibnamefont{Lu}}, \bibnamefont{and}
  \bibinfo{author}{\bibfnamefont{W.~J.} \bibnamefont{Schaff}},
  \bibinfo{journal}{Appl. Phys. Lett.} \textbf{\bibinfo{volume}{88}},
  \bibinfo{pages}{252109} (\bibinfo{year}{2006}).

\bibitem[{\citenamefont{Jones et~al.}(2007)\citenamefont{Jones, Li, Haller, van
  Genuchten, Yu, Ager, III, Liliental-Weber, Walukiewicz, Lu
  et~al.}}]{Jones2007}
\bibinfo{author}{\bibfnamefont{R.~E.} \bibnamefont{Jones}},
  \bibinfo{author}{\bibfnamefont{S.~X.} \bibnamefont{Li}},
  \bibinfo{author}{\bibfnamefont{E.~E.} \bibnamefont{Haller}},
  \bibinfo{author}{\bibfnamefont{H.~C.~M.} \bibnamefont{van Genuchten}},
  \bibinfo{author}{\bibfnamefont{K.~M.} \bibnamefont{Yu}},
  \bibinfo{author}{\bibfnamefont{J.~W.} \bibnamefont{Ager III}},
  \bibinfo{author}{\bibfnamefont{Z.}~\bibnamefont{Liliental-Weber}},
  \bibinfo{author}{\bibfnamefont{W.}~\bibnamefont{Walukiewicz}},
  \bibinfo{author}{\bibfnamefont{H.}~\bibnamefont{Lu}}, \bibnamefont{et~al.},
  \bibinfo{journal}{Appl. Phys. Lett.} \textbf{\bibinfo{volume}{90}},
  \bibinfo{eid}{162103} (\bibinfo{year}{2007}).

\bibitem[{\citenamefont{Schaff et~al.}(2004)\citenamefont{Schaff, Lu, Eastman,
  Walukiewicz, Yu, Keller, Kurtz, Keyes, and Gevilas}}]{Schaff2004}
\bibinfo{author}{\bibfnamefont{W.~J.} \bibnamefont{Schaff}},
  \bibinfo{author}{\bibfnamefont{H.}~\bibnamefont{Lu}},
  \bibinfo{author}{\bibfnamefont{L.~F.} \bibnamefont{Eastman}},
  \bibinfo{author}{\bibfnamefont{W.}~\bibnamefont{Walukiewicz}},
  \bibinfo{author}{\bibfnamefont{K.~M.} \bibnamefont{Yu}},
  \bibinfo{author}{\bibfnamefont{S.}~\bibnamefont{Keller}},
  \bibinfo{author}{\bibfnamefont{S.}~\bibnamefont{Kurtz}},
  \bibinfo{author}{\bibfnamefont{B.}~\bibnamefont{Keyes}}, \bibnamefont{and}
  \bibinfo{author}{\bibfnamefont{L.}~\bibnamefont{Gevilas}}, in
  \emph{\bibinfo{booktitle}{State-of-the-Art Program on Compound Semiconductors
  XLI and Nitride and Wide Bandgap Semiconductors for Sensors, Photonics, and
  Electronics V}}, edited by \bibinfo{editor}{\bibfnamefont{H.~M.}
  \bibnamefont{Ng}} \bibnamefont{and} \bibinfo{editor}{\bibfnamefont{A.~G.}
  \bibnamefont{Baca}} (\bibinfo{publisher}{Electrochemical Society},
  \bibinfo{address}{Honolulu, HI}, \bibinfo{year}{2004}), vol.
  \bibinfo{volume}{2004-06} of \emph{\bibinfo{series}{The Electrochemical
  Society Proceedings Series}}, p. \bibinfo{pages}{358}.

\bibitem[{\citenamefont{Arroyo Rojas~Dasilva et~al.}(2010)\citenamefont{Arroyo
  Rojas~Dasilva, Chauvat, Ruterana, Lahourcade, Monroy, and
  Nataf}}]{Dasilva2010}
\bibinfo{author}{\bibfnamefont{Y.}~\bibnamefont{Arroyo Rojas~Dasilva}},
  \bibinfo{author}{\bibfnamefont{M.~P.} \bibnamefont{Chauvat}},
  \bibinfo{author}{\bibfnamefont{P.}~\bibnamefont{Ruterana}},
  \bibinfo{author}{\bibfnamefont{L.}~\bibnamefont{Lahourcade}},
  \bibinfo{author}{\bibfnamefont{E.}~\bibnamefont{Monroy}}, \bibnamefont{and}
  \bibinfo{author}{\bibfnamefont{G.}~\bibnamefont{Nataf}}, \bibinfo{journal}{J.
  Phys. Cond. Matt.} \textbf{\bibinfo{volume}{22}}, \bibinfo{pages}{355802}
  (\bibinfo{year}{2010}).

\bibitem[{\citenamefont{Liliental-Weber
  et~al.}(2007)\citenamefont{Liliental-Weber, Jones, van Genuchten, Yu,
  Walukiewicz, III, Haller, Lu, and Schaff}}]{LilientalWeber2007}
\bibinfo{author}{\bibfnamefont{Z.}~\bibnamefont{Liliental-Weber}},
  \bibinfo{author}{\bibfnamefont{R.}~\bibnamefont{Jones}},
  \bibinfo{author}{\bibfnamefont{H.}~\bibnamefont{van Genuchten}},
  \bibinfo{author}{\bibfnamefont{K.}~\bibnamefont{Yu}},
  \bibinfo{author}{\bibfnamefont{W.}~\bibnamefont{Walukiewicz}},
  \bibinfo{author}{\bibfnamefont{J.~W.} \bibnamefont{Ager III}},
  \bibinfo{author}{\bibfnamefont{E.}~\bibnamefont{Haller}},
  \bibinfo{author}{\bibfnamefont{H.}~\bibnamefont{Lu}}, \bibnamefont{and}
  \bibinfo{author}{\bibfnamefont{W.}~\bibnamefont{Schaff}},
  \bibinfo{journal}{Physica B} \textbf{\bibinfo{volume}{401-402}},
  \bibinfo{pages}{646} (\bibinfo{year}{2007}).

\bibitem[{\citenamefont{Saarinen et~al.}(1998)\citenamefont{Saarinen,
  Hautoj\"{a}rvi, and Corbel}}]{Saarinen1998}
\bibinfo{author}{\bibfnamefont{K.}~\bibnamefont{Saarinen}},
  \bibinfo{author}{\bibfnamefont{P.}~\bibnamefont{Hautoj\"{a}rvi}},
  \bibnamefont{and} \bibinfo{author}{\bibfnamefont{C.}~\bibnamefont{Corbel}},
  \emph{\bibinfo{title}{Positron Annihilation Spectroscopy of Defects in
  Semiconductors}}, vol. \bibinfo{volume}{51A} of
  \emph{\bibinfo{series}{Semiconductors and Semimetals}}
  (\bibinfo{publisher}{Academic Press, New York}, \bibinfo{year}{1998}).

\bibitem[{\citenamefont{Rauch et~al.}(2011)\citenamefont{Rauch, Makkonen, and
  Tuomisto}}]{Rauch2011c}
\bibinfo{author}{\bibfnamefont{C.}~\bibnamefont{Rauch}},
  \bibinfo{author}{\bibfnamefont{I.}~\bibnamefont{Makkonen}}, \bibnamefont{and}
  \bibinfo{author}{\bibfnamefont{F.}~\bibnamefont{Tuomisto}},
  \bibinfo{journal}{Phys. Rev. B} \textbf{\bibinfo{volume}{84}},
  \bibinfo{pages}{125201} (\bibinfo{year}{2011}).

\bibitem[{\citenamefont{van Veen et~al.}(1995)\citenamefont{van Veen, Schut,
  Clement, de~Nijs, Kruseman, and IJpma}}]{Veen1995}
\bibinfo{author}{\bibfnamefont{A.}~\bibnamefont{van Veen}},
  \bibinfo{author}{\bibfnamefont{H.}~\bibnamefont{Schut}},
  \bibinfo{author}{\bibfnamefont{M.}~\bibnamefont{Clement}},
  \bibinfo{author}{\bibfnamefont{J.~M.~M.} \bibnamefont{de~Nijs}},
  \bibinfo{author}{\bibfnamefont{A.}~\bibnamefont{Kruseman}}, \bibnamefont{and}
  \bibinfo{author}{\bibfnamefont{M.~R.} \bibnamefont{IJpma}},
  \bibinfo{journal}{Appl. Surf. Sci.} \textbf{\bibinfo{volume}{85}},
  \bibinfo{pages}{216} (\bibinfo{year}{1995}).

\bibitem[{\citenamefont{Tuomisto
  et~al.}(2007{\natexlab{a}})\citenamefont{Tuomisto, Pelli, Yu, Walukiewicz,
  and Schaff}}]{Tuomisto2007e}
\bibinfo{author}{\bibfnamefont{F.}~\bibnamefont{Tuomisto}},
  \bibinfo{author}{\bibfnamefont{A.}~\bibnamefont{Pelli}},
  \bibinfo{author}{\bibfnamefont{K.~M.} \bibnamefont{Yu}},
  \bibinfo{author}{\bibfnamefont{W.}~\bibnamefont{Walukiewicz}},
  \bibnamefont{and} \bibinfo{author}{\bibfnamefont{W.~J.}
  \bibnamefont{Schaff}}, \bibinfo{journal}{Phys. Rev. B}
  \textbf{\bibinfo{volume}{75}}, \bibinfo{pages}{193201}
  (\bibinfo{year}{2007}{\natexlab{a}}).

\bibitem[{\citenamefont{Reurings et~al.}(2010)\citenamefont{Reurings, Rauch,
  Tuomisto, Jones, Yu, Walukiewicz, and Schaff}}]{Reurings2010}
\bibinfo{author}{\bibfnamefont{F.}~\bibnamefont{Reurings}},
  \bibinfo{author}{\bibfnamefont{C.}~\bibnamefont{Rauch}},
  \bibinfo{author}{\bibfnamefont{F.}~\bibnamefont{Tuomisto}},
  \bibinfo{author}{\bibfnamefont{R.~E.} \bibnamefont{Jones}},
  \bibinfo{author}{\bibfnamefont{K.~M.} \bibnamefont{Yu}},
  \bibinfo{author}{\bibfnamefont{W.}~\bibnamefont{Walukiewicz}},
  \bibnamefont{and} \bibinfo{author}{\bibfnamefont{W.~J.}
  \bibnamefont{Schaff}}, \bibinfo{journal}{Phys. Rev. B}
  \textbf{\bibinfo{volume}{82}}, \bibinfo{pages}{153202}
  (\bibinfo{year}{2010}).

\bibitem[{\citenamefont{Limpijumnong and Van~de
  Walle}(2004)}]{Limpijumnong2004}
\bibinfo{author}{\bibfnamefont{S.}~\bibnamefont{Limpijumnong}}
  \bibnamefont{and} \bibinfo{author}{\bibfnamefont{C.~G.} \bibnamefont{Van~de
  Walle}}, \bibinfo{journal}{Phys. Rev. B} \textbf{\bibinfo{volume}{69}},
  \bibinfo{pages}{035207} (\bibinfo{year}{2004}).

\bibitem[{\citenamefont{Duan and Stampfl}(2009)}]{Duan2009a}
\bibinfo{author}{\bibfnamefont{X.~M.} \bibnamefont{Duan}} \bibnamefont{and}
  \bibinfo{author}{\bibfnamefont{C.}~\bibnamefont{Stampfl}},
  \bibinfo{journal}{Phys. Rev. B} \textbf{\bibinfo{volume}{79}},
  \bibinfo{pages}{174202} (\bibinfo{year}{2009}).

\bibitem[{\citenamefont{Tuomisto
  et~al.}(2007{\natexlab{b}})\citenamefont{Tuomisto, Paskova, Kroger, Figge,
  Hommel, Monemar, and Kersting}}]{Tuomisto2007c}
\bibinfo{author}{\bibfnamefont{F.}~\bibnamefont{Tuomisto}},
  \bibinfo{author}{\bibfnamefont{T.}~\bibnamefont{Paskova}},
  \bibinfo{author}{\bibfnamefont{R.}~\bibnamefont{Kroger}},
  \bibinfo{author}{\bibfnamefont{S.}~\bibnamefont{Figge}},
  \bibinfo{author}{\bibfnamefont{D.}~\bibnamefont{Hommel}},
  \bibinfo{author}{\bibfnamefont{B.}~\bibnamefont{Monemar}}, \bibnamefont{and}
  \bibinfo{author}{\bibfnamefont{R.}~\bibnamefont{Kersting}},
  \bibinfo{journal}{Appl. Phys. Lett.} \textbf{\bibinfo{volume}{90}},
  \bibinfo{pages}{121915} (\bibinfo{year}{2007}{\natexlab{b}}).

\bibitem[{\citenamefont{M\"aki et~al.}(2011)\citenamefont{M\"aki, Makkonen,
  Tuomisto, Karjalainen, Suihkonen, R\"ais\"anen, Chemekova, and
  Makarov}}]{Maki2011}
\bibinfo{author}{\bibfnamefont{J.-M.} \bibnamefont{M\"aki}},
  \bibinfo{author}{\bibfnamefont{I.}~\bibnamefont{Makkonen}},
  \bibinfo{author}{\bibfnamefont{F.}~\bibnamefont{Tuomisto}},
  \bibinfo{author}{\bibfnamefont{A.}~\bibnamefont{Karjalainen}},
  \bibinfo{author}{\bibfnamefont{S.}~\bibnamefont{Suihkonen}},
  \bibinfo{author}{\bibfnamefont{J.}~\bibnamefont{R\"ais\"anen}},
  \bibinfo{author}{\bibfnamefont{T.~Y.} \bibnamefont{Chemekova}},
  \bibnamefont{and} \bibinfo{author}{\bibfnamefont{Y.~N.}
  \bibnamefont{Makarov}}, \bibinfo{journal}{Phys. Rev. B}
  \textbf{\bibinfo{volume}{84}}, \bibinfo{pages}{081204}
  (\bibinfo{year}{2011}).

\bibitem[{\citenamefont{Duan and Stampfl}(2008)}]{Duan2008}
\bibinfo{author}{\bibfnamefont{X.~M.} \bibnamefont{Duan}} \bibnamefont{and}
  \bibinfo{author}{\bibfnamefont{C.}~\bibnamefont{Stampfl}},
  \bibinfo{journal}{Phys. Rev. B} \textbf{\bibinfo{volume}{77}},
  \bibinfo{eid}{115207} (\bibinfo{year}{2008}).

\bibitem[{\citenamefont{Uedono et~al.}({2009})\citenamefont{Uedono, Nakamori,
  Narita, Suzuki, Wang, Che, Ishitani, Yoshikawa, and Ishibashi}}]{Uedono2009b}
\bibinfo{author}{\bibfnamefont{A.}~\bibnamefont{Uedono}},
  \bibinfo{author}{\bibfnamefont{H.}~\bibnamefont{Nakamori}},
  \bibinfo{author}{\bibfnamefont{K.}~\bibnamefont{Narita}},
  \bibinfo{author}{\bibfnamefont{J.}~\bibnamefont{Suzuki}},
  \bibinfo{author}{\bibfnamefont{X.}~\bibnamefont{Wang}},
  \bibinfo{author}{\bibfnamefont{S.~B.} \bibnamefont{Che}},
  \bibinfo{author}{\bibfnamefont{Y.}~\bibnamefont{Ishitani}},
  \bibinfo{author}{\bibfnamefont{A.}~\bibnamefont{Yoshikawa}},
  \bibnamefont{and}
  \bibinfo{author}{\bibfnamefont{S.}~\bibnamefont{Ishibashi}},
  \bibinfo{journal}{J. Appl. Phys.} \textbf{\bibinfo{volume}{{105}}},
  \bibinfo{pages}{054507} (\bibinfo{year}{{2009}}).

\bibitem[{\citenamefont{Kraeusel and Hourahine}(unpublished)}]{Kraeusel2011}
\bibinfo{author}{\bibfnamefont{S.}~\bibnamefont{Kraeusel}} \bibnamefont{and}
  \bibinfo{author}{\bibfnamefont{B.}~\bibnamefont{Hourahine}}
  \bibinfo{journal}{Phys. Status Solidi A} \bibinfo{doi}{DOI:10.1002/pssa.201100097} (\bibinfo{year}{2011}).

\bibitem[{\citenamefont{Oila et~al.}(2003)\citenamefont{Oila, Kivioja, Ranki,
  Saarinen, Look, Molnar, Park, Lee, and Han}}]{Oila2003}
\bibinfo{author}{\bibfnamefont{J.}~\bibnamefont{Oila}},
  \bibinfo{author}{\bibfnamefont{J.}~\bibnamefont{Kivioja}},
  \bibinfo{author}{\bibfnamefont{V.}~\bibnamefont{Ranki}},
  \bibinfo{author}{\bibfnamefont{K.}~\bibnamefont{Saarinen}},
  \bibinfo{author}{\bibfnamefont{D.~C.} \bibnamefont{Look}},
  \bibinfo{author}{\bibfnamefont{R.~J.} \bibnamefont{Molnar}},
  \bibinfo{author}{\bibfnamefont{S.~S.} \bibnamefont{Park}},
  \bibinfo{author}{\bibfnamefont{S.~K.} \bibnamefont{Lee}}, \bibnamefont{and}
  \bibinfo{author}{\bibfnamefont{J.~Y.} \bibnamefont{Han}},
  \bibinfo{journal}{Appl. Phys. Lett.} \textbf{\bibinfo{volume}{82}},
  \bibinfo{pages}{3433} (\bibinfo{year}{2003}).

\bibitem[{\citenamefont{Tengborn et~al.}(2006)\citenamefont{Tengborn,
  Rummukainen, Tuomisto, Saarinen, Rudzinski, Hageman, Larsen, and
  Nordlund}}]{Tengborn2006}
\bibinfo{author}{\bibfnamefont{E.}~\bibnamefont{Tengborn}},
  \bibinfo{author}{\bibfnamefont{M.}~\bibnamefont{Rummukainen}},
  \bibinfo{author}{\bibfnamefont{F.}~\bibnamefont{Tuomisto}},
  \bibinfo{author}{\bibfnamefont{K.}~\bibnamefont{Saarinen}},
  \bibinfo{author}{\bibfnamefont{M.}~\bibnamefont{Rudzinski}},
  \bibinfo{author}{\bibfnamefont{P.~R.} \bibnamefont{Hageman}},
  \bibinfo{author}{\bibfnamefont{P.~K.} \bibnamefont{Larsen}},
  \bibnamefont{and} \bibinfo{author}{\bibfnamefont{A.}~\bibnamefont{Nordlund}},
  \bibinfo{journal}{Appl. Phys. Lett.} \textbf{\bibinfo{volume}{89}},
  \bibinfo{pages}{091905} (\bibinfo{year}{2006}).

\end{thebibliography}
\end{document}